\documentclass[aps,prl,twocolumn,superscriptaddress,floatfix]{revtex4-2}
\usepackage[colorlinks=true,linkcolor=blue,urlcolor=blue,citecolor=blue,pdfusetitle]{hyperref}
\usepackage[utf8]{inputenc}
\usepackage[english]{babel}
\usepackage{amsmath}
\usepackage[caption = false]{subfig}
\usepackage{graphicx,epstopdf}
\usepackage{blindtext}
\usepackage[table,xcdraw]{xcolor}
\usepackage{lipsum}
\usepackage{amsfonts}
\usepackage{bbm}
\usepackage{amssymb}
\usepackage{enumerate}
\usepackage{color}
\usepackage{latexsym}
\usepackage{physics}
\usepackage{times,txfonts}
\usepackage{pdfcomment}

\newcommand{\SubFig}[2]{\ref{#1}{\color{blue}#2}}

\definecolor{MyGreen}{RGB}{0, 179, 134}
\definecolor{MyRed}{RGB}{255, 102, 102}

\newcommand{\CSIC}{\href{https://ror.org/009wseg80}{Instituto de Física Fundamental}, \href{https://ror.org/02gfc7t72}{Consejo Superior de Investigaciones Científicas}, Calle Serrano 113b, Madrid 28006, Spain}

\usepackage{orcidlink}

\begin{document}

\title{Decoherence-Free Qubit and Chiral Emission from a Giant Molecule in Waveguide QED}

\author{Yang Wang\orcidlink{0009-0000-6778-5285}}
\email{yang.wang01@estudiante.uam.es}
\affiliation{\CSIC}

\author{Juan José García-Ripoll\orcidlink{0000-0001-8993-4624}}
\email{jj.garcia.ripoll@csic.es}
\affiliation{\CSIC}

\author{Alan C. Santos\orcidlink{0000-0002-6989-7958}}
\email{ac\_santos@iff.csic.es}
\affiliation{\CSIC}

\begin{abstract}
	Combining decoherence protection with directional photon emission in a single waveguide quantum electrodynamics (QED) device remains an open challenge. Here we show that an artificial giant molecule---strongly interacting artificial atoms coupled to a photonic waveguide at multiple spatially separated points---achieves both: a fully operational decoherence-free (DF) qubit and state-dependent chiral single-photon emission, arising from the same photon-interference mechanism. Initialization reduces to a local excitation of a single atom, universal single-qubit gates are implemented by modulating a single atomic frequency, and readout exploits state-dependent chiral emission with directionality reaching 100\% and low measurement error of 1.2\%. The coexistence of decoherence protection and directional emission in a single device positions giant molecules as protected chiral nodes for modular quantum networks in waveguide QED.
\end{abstract}

\maketitle


\paragraph{Introduction.}

Waveguide quantum electrodynamics (QED) has emerged as a powerful platform for quantum information processing, where artificial atoms coupled to propagating photons enable long-range and programmable interactions between distant emitters~\cite{sheremet:2023,niu:2023,croot:2025,gu:2017,blais:2021,leung:2019}. However, the same coupling to guided modes that mediates coherent interactions also leads to strong radiative decay into the waveguide, which acts as an additional source of decoherence. Protecting quantum information from this dissipation therefore represents a central challenge for waveguide QED architectures. One promising strategy relies on engineering symmetry-protected states belonging to decoherence-free (DF) subspaces~\cite{gutierrez-jauregui:2023,karnieli:2025,paulisch:2016,qiao:2022,holzinger:2022,rubies-bigorda:2025}. In parallel, the recently introduced paradigm of giant atoms—artificial atoms coupled to a waveguide at multiple spatially separated points—offers new possibilities to engineer DF interactions through interference between different emission pathways~\cite{kockum:2018}. Significant experimental progress toward these ideas has already been achieved in superconducting circuit platforms~\cite{kannan:2020,zanner:2022}.

Despite these advances, practical DF qubits in waveguide QED have yet to be demonstrated. Although several schemes for creating DF states have been proposed~\cite{paulisch:2016,qiao:2022,holzinger:2022,rubies-bigorda:2025}, implementing the basic requirements of a physical qubit—initialization, coherent control, and readout of fiducial states~\cite{divincenzo:2000}—typically involves complex many-body operations. In particular, existing approaches often rely on collective Dicke states~\cite{wang:2020,zanner:2022,sheremet:2023,vandiepen:2025}, which encode information in highly entangled states of atoms distributed along the waveguide. The nonlocal and many-body nature of these states makes their preparation, manipulation, and measurement experimentally demanding~\cite{pineiroorioli:2022}. Developing experimentally feasible DF qubit encodings that allow simple initialization and efficient readout therefore remains an important open challenge.

A second central challenge is achieving state-dependent directional photon emission, a prerequisite for routing quantum information between distant nodes in modular architectures~\cite{lodahl:2017,kannan:2023,almanakly:2025}. Recent experiments have demonstrated on-demand chiral emission from artificial atoms~\cite{kannan:2023,joshi:2023} and chiral quantum interconnects between superconducting modules~\cite{almanakly:2025}, yet no single device has so far combined decoherence protection with directional emission. Notably, the interference physics underlying giant-atom decoherence-free interactions is closely related to the mechanisms that produce chirality~\cite{soro:2022,roccati:2024,du:2025}, suggesting that both capabilities can be unified.

\begin{figure}[t!]
	\centering
	\includegraphics[width=\linewidth]{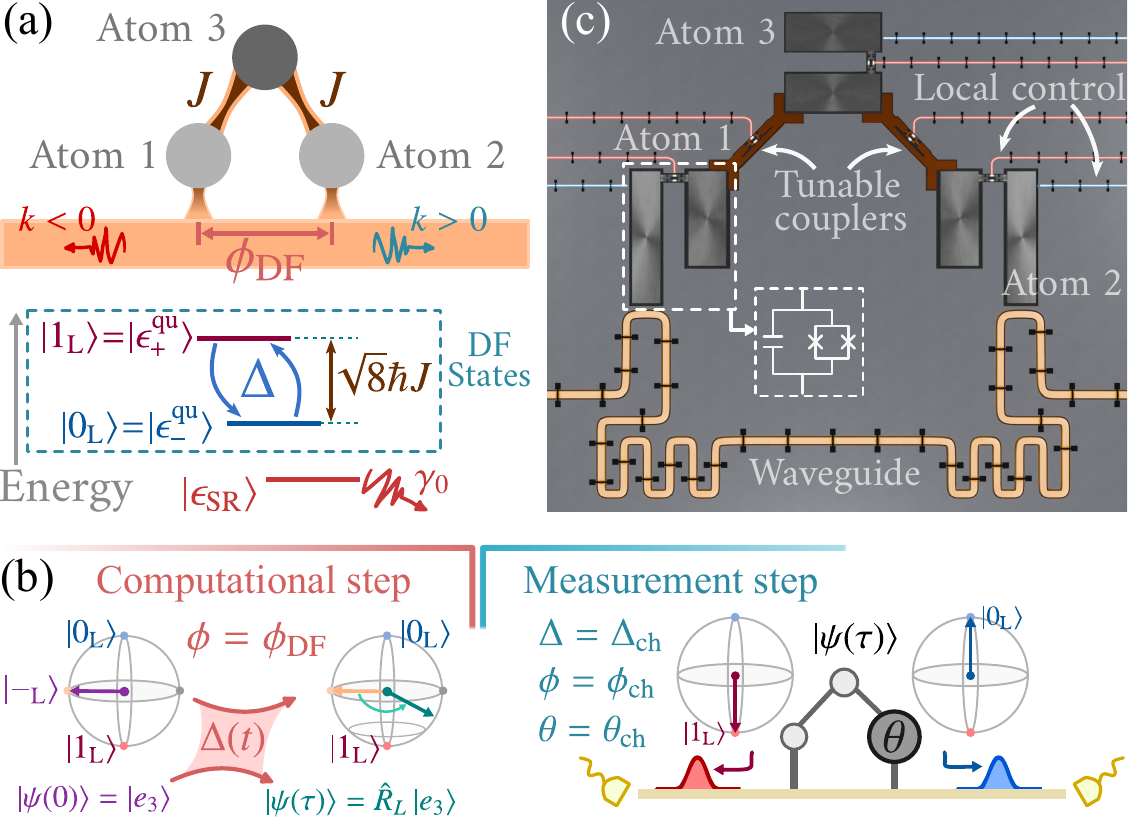}
	\caption{(a) Giant triatomic molecule coupled to a photonic waveguide. At phase $\phi_{\mathrm{DF}}=\pi$, the molecule supports two decoherence-free logical qubit states $\ket{0_L}$ and $\ket{1_L}$ and a super-radiant state. The tunable detuning $\Delta$ of atom 3 drives transitions and universal gates within the DF subspace.
	(b) Superconducting-circuit implementation. Tunable couplers mediate intramolecular interactions, microwave XY pulses initialize atom 3, and dedicated flux lines modulate atomic frequencies for single-qubit gates and chiral readout.
	(c) A local phase-shift gate $\hat{S}(\theta)$ on atom 2, combined with tuning $\phi$ and $\Delta$, enables state-dependent directional emission and projective measurements.}
	\label{Fig1:System_LogicGates}
\end{figure}

In this paper we introduce a waveguide QED based symmetry-protected qubit that supports full control (initialization, arbitrary single-qubit unitaries) and high-fidelity readout via chiral emission. The device, depicted in Fig.~\SubFig{Fig1:System_LogicGates}{a}, is a triatomic artificial giant molecule coupled to a bidirectional waveguide at two distant points. Photon interference at the spatially separated coupling points~\cite{kockum:2018} gives rise to two fully controllable dark states in the single-excitation manifold, which serve as the logical qubit basis, and simultaneously enables 100\% directional emission from states prepared through local controls. This state-dependent chirality provides projective measurements in the logical basis (see Fig.~\SubFig{Fig1:System_LogicGates}{b}), without requiring nonlocal control or many-body state tomography.


\paragraph{Triatomic Giant Molecule.}

A giant molecule couples to the waveguide at two spatially separated points, so that the photon phase $\phi$ accumulated between the coupling points governs both the decoherence properties and the directionality of emission. Our realization consists of three interacting artificial atoms coupled to the continuum of modes of a 1D photonic waveguide, constituting a single \textit{artificial molecule} in waveguide QED.

Specifically, as depicted in Fig.~\ref{Fig1:System_LogicGates}, the molecule is composed of two-level atoms with frequencies $\omega_{\mathrm{at},n}$. The atoms labeled 1 and 2, the guided atoms, are directly connected to the waveguide. The non-guided atom 3 only interacts with the atoms 1 and 2, with coupling strength $J_{n}$ satisfying $|J_{n}| \ll |\omega_{\mathrm{at},n}|$. The guided atoms interact with the waveguide modes with a coupling function $V_{k,n} =g_{n}(k)e^{ikx_{n}}$, with $|g_{n}(k)|$ as the interaction strength of the atom with a $k$-th mode in the waveguide, and $x_{n}$ the position of the $n$-th atom. The coupling strength of the atoms with left- and right-propagating photons are identical, i.e. $g_{n}(k>0)=g_{n}(k<0)$. The Hamiltonian of the system is
\begin{equation}
	\hat{H}_{\mathrm{total}}= \hat{H}_{\mathrm{at}}+\hat{H}_{\mathrm{f}}+\hat{H}_{\mathrm{mol}} +\hat{H}_{\mathrm{mw}} , \label{Eq:TotalH}
\end{equation}
where $\hat{H}_{\mathrm{at}}=\hbar \sum_{n} \omega_{\mathrm{at},n}\hat{\sigma}_{n}^{+}\hat{\sigma}_{n}^{-}$ and $\hat{H}_{\mathrm{f}} = \int_{-\infty}^{+\infty}dk \hbar \omega_{k} \hat{a}_{k}^{\dagger}\hat{a}_{k}$ describe the self energies of atoms and the electromagnetic field in the waveguide, respectively, and
\begin{align}
	\hat{H}_{\mathrm{mol}} &= \hbar\sum \nolimits _{n=\{1,2\}} J_{n}(\hat{\sigma}_{n}^{+}\hat{\sigma}_{3}^{-} + \hat{\sigma}_{n}^{-}\hat{\sigma}_{3}^{+}) , \label{Eq:Hmol} \\
	\hat{H}_{\mathrm{mw}} &= \hbar\sum_{n=\{1,2\}}\int_{-\infty}^{+\infty}dk \left( V_{k,n}^{*} \hat{a}_{k}^{\dagger} \hat{\sigma}_{n}^{-}+V_{k,n}\hat{\sigma}_{n}^{+}\hat{a}_{k} \right) , \label{Eq:Hef}
\end{align}
are the molecule interaction Hamiltonian and molecule-waveguide coupling, respectively, with $\hat{\sigma}_{n}^{+}$ and $\hat{\sigma}_{n}^{-}$ the raising and lowering operators on atom $n$, and $\hat{a}_{k}^{\dagger} $ ($\hat{a}_{k} $) is the photon creation (annihilation) operator with wavenumber $k$.

The giant nature of the molecule arises from its coupling to the waveguide at multiple points, namely $x_{1}$ and $x_{2}$, and the hybridization of the atomic states in the molecule. The accumulated phase $\phi_{k} = k|x_{2}-x_{1}|$ for a photon, with wavenumber $k$, leads to interference effects that cannot be explained with a multi-level electronic molecule interacting with the electromagnetic field under the dipole approximation, similarly to giant atoms theory~\cite{kockum:2018,kannan:2020,du:2025,levy-yeyati:2025}.

Throughout this manuscript we assume identical intra-molecular interactions, $J_{n}=J$, atomic frequencies $\omega_{\mathrm{at},n=\{1,2\}} = \omega_{0}$, with a possible detuning on the third atom $\Delta := \omega_{\mathrm{at},3} - \omega_{0}$. We also introduce the wavenumber $k_0$ that is resonant with the atomic frequency $\omega_0$ and define $\phi =  \phi_{k_{0}} = k_{0}|x_{2}-x_{1}|$. Note that this phase can be tuned (modulo $2\pi$) by simultaneously adjusting the frequencies of all three qubits $\omega_0$.

In the single-excitation sector, the Wigner-Weisskopf wavefunction~\cite{garciaripoll:2022} describes the joint molecule-waveguide dynamics
\begin{equation}
	\ket{\Psi(t)}= e^{-i\omega_0t}\left[ \ket{G}\int_{-\infty}^{+\infty} \alpha_{k}(t) \hat{a}_{k}^{\dagger} dk+\sum_{n=1}^{3}c_{n}(t)\ket{e_n}  \right]\ket{\emptyset}. \label{Eq:Ansatz}
\end{equation}
in terms of the molecular ground state $\ket{G}$, the photonic vacuum $\ket{\emptyset}$, the atomic excited states $\ket{e_n}:=\hat{\sigma}_{n}^{+}\ket{G}$, and the photon and atom amplitudes $\alpha_{k}(t)$ and $c_{n}(t)$.

Under the Born-Markov approximation~\cite{breuer:2007,walls:2008}, standard in giant-atom theory~\cite{kockum:2018,kannan:2020}, the unnormalized molecular state $\ket{c(t)} = \sum_{n=1}^{3}c_{n}(t)\ket{e_n}$ obeys $i\hbar \ket{\dot{c}(t)} = \hat{\mathcal{H}}\ket{c(t)}$, with the non-Hermitian effective Hamiltonian
\begin{align}
	\hat{\mathcal{H}}
	& = \frac{\hbar}{2}
	\begin{bmatrix}
		-i\gamma_{0} & -i\gamma_{0}e^{i\phi} & 2 J \\
		-i\gamma_{0}e^{i\phi} & -i\gamma_{0} & 2 J \\
		2J & 2J & 2\Delta
	\end{bmatrix}.
	\label{Eq:SingleExcHamiltonian}
\end{align}
is defined by the intramolecular interaction $J$, the photon phase $\phi$, and the waveguide-induced decay rate $\gamma_{0}$.

The molecule's dynamics and emission properties derive from the operator's spectral decomposition
\begin{align}
	\hat{\mathcal{H}} &=
	\epsilon_0 \ketbra{\epsilon_0}{\epsilon_0'} + \epsilon_+ \ketbra{\epsilon_+}{\epsilon_+'}
	+ \epsilon_- \ketbra{\epsilon_-}{\epsilon_-'},\\
	\epsilon_{0}&=\frac{i \hbar}{2}\left( e^{i\phi}  - 1\right) \gamma_{0} ,~~~~
	\epsilon_{\pm}=\frac{\hbar}{4}\left[ 2\Delta - i\left(1+e^{i\phi}\right)\gamma_{0} \pm \Gamma \right].
	\label{Eq:Eigenvalues}
\end{align}
Here, $\Gamma^2= 4\big(\Delta ^2+8 J^2\big) - \gamma _0 \big(1+e^{i \phi }\big) \big[\gamma _0 (1+e^{i \phi })-4 i \Delta \big]$, and we have introduced the left and right eigenvectors with their orthogonality relation $\braket{\epsilon'_r}{\epsilon_s}=\delta_{rs}$. The eigenvalues' real and imaginary parts $\epsilon_{r}=\hbar\omega_r -i\hbar\gamma_r$ determine the eigenenergies $\omega_r$ and the decay rates $\gamma_r$ of the three eigenstates.

\paragraph{Symmetry-protected logical qubit.} The giant molecule supports a protected logical qubit encoded in two W-like states~\cite{agrawal:2006,jung:2008}
\begin{align}
	\ket{\epsilon_{\pm}} =  \frac{1}{2}\left( \ket{e_{1}}+\ket{e_{2}}  \pm \sqrt{2}\ket{e_{3}} \right),\label{Eq:LogicBasis}
\end{align}
with logical basis $\ket{1_{\mathrm{L}}}:=\ket*{\epsilon_{+}^{\mathrm{qu}}}$ and $\ket{0_{\mathrm{L}}}:=\ket*{\epsilon_{-}^{\mathrm{qu}}}$, defined entirely within the single-excitation manifold without participation of the molecule's ground state $\ket{G}$. These states become exactly decoherence-free at resonant phase $\phi=\phi_\text{DF}:=\pi$ and zero detuning $\Delta=0$, where the eigenvalues $\epsilon_{\pm}=\pm\sqrt{2}\hbar J$ are purely real. The triatomic molecule is the minimal giant-molecule geometry supporting a two-dimensional DF subspace.


\paragraph{Full controllability of the logical qubit.}
In our device, the fiducial state used for qubit initialization is a localized excitation on atom 3. Such input state $\ket{e_{3}}$ is a superposition of the logical qubit states $\ket{e_3} = \ket*{-_{L}} = (\ket{1}_{L} - \ket{0}_{L})/\sqrt{2}$ and can be prepared locally on each atom. This avoids the multi-qubit or non-local operations demanded by other protocols and setups.

In addition to qubit initialization, we can implement arbitrary single-qubit operations in the protected space through a time-dependent modulation of the detuning $\Delta(t)$. Note that the logical qubit follows a Hamiltonian evolution, governed by
\begin{equation}
	\hat{\mathcal{H}}_{\mathrm{qu}} = \sqrt{2} \hbar J\hat{\sigma}^{L}_{z} -\frac{1}{2} \hbar \Delta (t) \hat{\sigma}^{L}_{x} , \label{Eq:Hqubit}
\end{equation}
with $\hat{\sigma}^{L}_{z} = \ket{1_{\mathrm{L}}}\bra{1_{\mathrm{L}}} - \ket{0_{\mathrm{L}}}\bra{0_{\mathrm{L}}}$ and $\hat{\sigma}^{L}_{x} = \ket{1_{\mathrm{L}}}\bra{0_{\mathrm{L}}} + \ket{0_{\mathrm{L}}}\bra{1_{\mathrm{L}}}$ the logical Pauli matrices. We now assume a monochromatic parametric modulation $\Delta(t) = \Delta_{0} \cos( \omega_{\mathrm{d}} t + \varphi)$ of atom 3, which can be implemented in superconducting circuits with local flux lines acting on the third atom~\cite{roth:2017,wu:2018,hu:2026} (see Fig.~\SubFig{Fig1:System_LogicGates}{b}). This modulation leads to an effective Hamiltonian in the rotating frame defined by the coupling $J$
\begin{equation}
	\hat{\mathcal{H}}^{\mathrm{eff}}_{\mathrm{qu}} = \frac{\hbar(2\sqrt{2}J - \omega_d)}{2}  \hat{\sigma}_{z}^{L} - \frac{\hbar\Delta_{0}}{4}\left(\hat{\sigma}_{x}^{L} \cos\varphi +  \hat{\sigma}_{y}^{L} \sin \varphi\right).
\end{equation}
This is a universal Hamiltonian that can implement arbitrary rotations in the Bloch sphere~\cite{nielsen:2011} by a suitable control of the monochromatic drive $\omega_d$, the driving amplitude $\Delta_0$, the phase $\varphi$ and the parametric drive's duration.

Note that many single-qubit gates do not require a parametric modulation of the detuning. Phase gates are obtained by the trivial evolution with Hamiltonian~\eqref{Eq:Hqubit} under no detuning $\Delta=0$. Hadamard gates generated by $\hat{H} = \hbar \Delta_{0} (\hat{\sigma}_{z}+\hat{\sigma}_{x})$~\cite{santos:2018a} are obtained with constant detuning $\Delta(t) = \Delta_{0} = - 2\sqrt{2}J$ acting on the logical qubit over a time $\tau_{H} = \pi / 2J$.

Let us remark that we have demonstrated full initialization and operation with local control at the level of artificial atoms. Our platform does not require multi-qubit entangling operations~\cite{paulisch:2016}, transitions through short-lived super-radiant states that introduce errors during gate operations~\cite{rubies-bigorda:2025}, or global control over the full collective Hilbert space for initialization and, as seen now, readout of collective dark states~\cite{wang:2020,zanner:2022}.


\begin{figure}[t!]
	\centering
	\includegraphics[width=\linewidth]{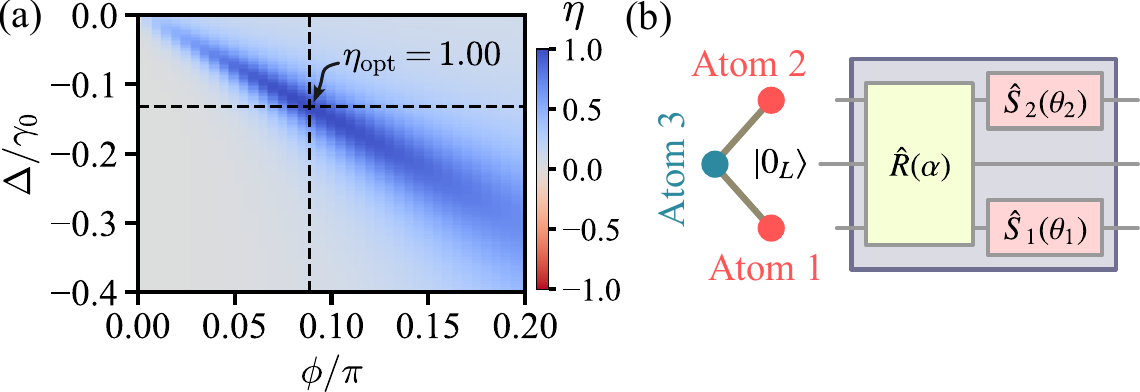}
	\caption{(a) Maximum chirality $\eta$ over all single-excitation molecular states $\ket{c(0)}$, as a function of $\Delta$ and $\phi$, for $J=0.1\gamma_0$. (b) Quantum circuit used to prepare the maximally chiral state $\ket{c_+}$ with $\eta=+1$.}
	\label{Fig:Directionality}
\end{figure}

\paragraph{Perfect chiral emission.} Directional single-photon emission is a key resource for quantum information routing in waveguide QED~\cite{kannan:2023,almanakly:2025}. The giant molecule supports perfect left- and right-wards chiral emission, enabled by the same interference mechanism that enables the protected qubit subspace, using states that can be created from the protected qubit space. State-dependent chiral emission can therefore be used to perform projective measurements of the logical basis.

The giant molecule emits energy in the form of right- and left-moving single photon wavepackets, $\alpha_R$ and $\alpha_L$, given by
\begin{equation}
	\alpha_{L(R)}(t)=\sqrt{\frac{\gamma_{0}}{2}}\left[ c_{1}(t)+e^{+(-)i\phi}c_{2}(t) \right]
	=: \braket{a_{L(R)}}{c(t)}
	\label{Eq:Input-Output-Field}
\end{equation}
in the Wigner-Weisskopf subspace.
The integrated chiral currents are a quadratic form of the molecule's initial state
\begin{equation}
	I_{L(R)} := \int_0^{+\infty} \left|\braket{\alpha_{L(R)}}{c(t)}\right|^2\mathrm{d}t =
	\matrixel{c(0)}{\hat{I}_{L(R)}}{c(0)},
\end{equation}
with operators $\hat{I}_{L,R}$ derived from the eigenstate decomposition of $\mathcal{H}$. The chirality
\begin{equation}
	\eta = \frac{I_{R} - I_{L}}{I_{R} + I_{L}}  \in [-1,1] ,
\end{equation}
quantifies the directionality, with $\eta=+1$ and $\eta=-1$ corresponding to perfect right- and leftwards emission. These extreme values are achieved by initial states satisfying
\begin{equation}
	(\hat{I}_R-\hat{I}_L)\ket{c_\eta} = \eta (\hat{I}_R+\hat{I}_L)\ket{c_\eta},~\mbox{with}~\eta=\pm 1.
\end{equation}

Perfect 100\% rightwards chiral emission is possible outside the protected qubit regime (see Fig.~\SubFig{Fig:Directionality}{a}). It can be found by exploring values of $(\gamma,J,\Delta,\phi)$ to find the $\{\hat{I}_R\pm\hat{I}_L\}$ operators with generalized eigenvalues $\eta=\pm 1$. For $J=0.1\gamma_0$, we must tune the reference frequency $\omega_0$ to achieve $\phi_\text{ch}=0.088\pi~\mathrm{mod}~2\pi$ and set $\Delta_\text{ch}=-0.132\gamma_0$. The ideal state $\ket{c_+}$ that only emits photons to the right can be constructed from the logical qubit's zero state as
\begin{equation}
\ket{c_+} = \prod_{n=1,2} S_n(\theta_n) R(\alpha) \ket{0_L} =: U_+\ket{0_L}.
\label{Eq:PerfectChiralStates}
\end{equation}
This involves a logical qubit gate $\hat{R}(\alpha) = \exp(- i \hat{\sigma}_y^L \alpha)$, and two local gates $\hat{S}_{n}(\theta_{n})$ acting on atoms 1 and 2, with angles $(\theta_1,\theta_2,\alpha)\simeq (-0.0443\pi, 0.8671\pi, 0.0031\pi)$ (cf. Fig.~\SubFig{Fig:Directionality}{b}).

\begin{figure}
	\centering
	\includegraphics[width=\linewidth]{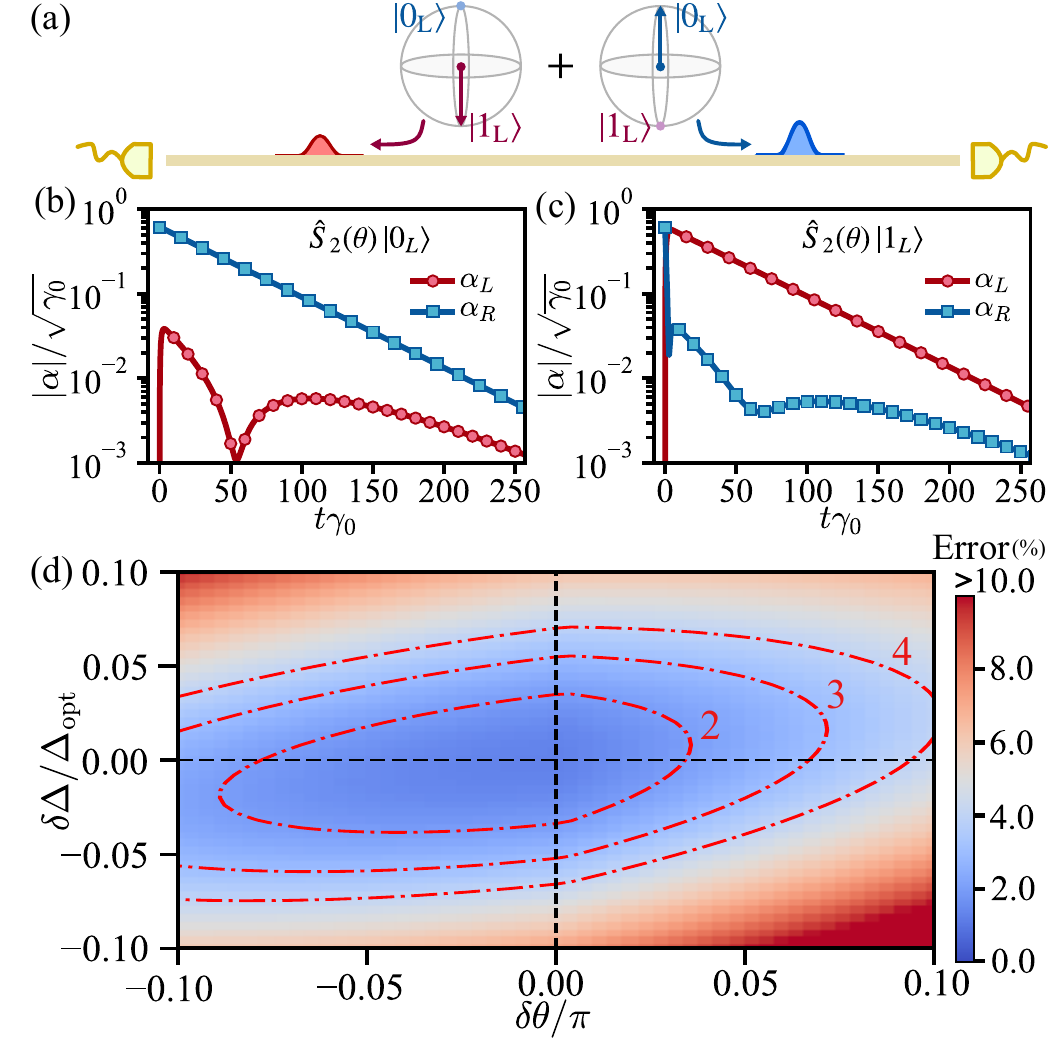}
	\caption{(a) Measurement protocol based on chiral emission from the $\ket{0_L}$ and $\ket{1_L}$ states. A photon detected on the right (left) port is assigned to a projective measurement of the $\ket{0_L}$ ($\ket{1_L}$) state. Measurement errors are dominated by imperfect chiral emission from the $\ket{1_L}$ state. (b,c) Photon field amplitudes $\alpha_{L}(t)$ and $\alpha_{R}(t)$ for the states (b) $S_2(\theta)\ket{0_L}$ and (c) $S_2(\theta)\ket{1_L}$, at the point of maximum chirality $(J,\Delta_\text{ch},\phi_\text{ch})\simeq(0.1\gamma_0,-0.132\gamma_0,0.088\pi)$. (d) Maximum measurement error for logical qubit states as a function of the deviations $\delta \theta$ and $\delta \Delta$ from the second measurement protocol values $(\Delta,\phi,\theta) \simeq (-0.1293\gamma_{0},0.0266\pi,0.9974\pi)$. At the optimal point ($\delta \theta=\delta \Delta=0$), the maximum error is $1.27\%$. Dash-dotted curves enclose the regions with maximum measurement error 2\%, 3\% and 4\%. All panels use $\phi=\phi_{\mathrm{ch}}$ and $J=0.1\gamma_0$.}
	\label{Fig:MeasuError}
\end{figure}
\paragraph{Projective measurements from chiral emission.}
The chiral emission process defines a family of projective measurements on the logical qubit space. The simplest protocol is based on Eq.~\eqref{Eq:PerfectChiralStates}. In this protocol, we first act on the logical qubit state with the transformation $U_+$ and then change $(\Delta,\phi)$ to enter the chiral emission regime. Our measurement assigns $\ket{0_L}$ and $\ket{1_L}$ when a photon is detected on the right and left ports, respectively (cf. Fig.~\SubFig{Fig:MeasuError}{a}).

We define the detection error of the logical state $\ket{0_{L}}$ as the probability of emitting a leftward-propagating photon when the system is initialized in $\ket{0_{L}}$, i.e., $E_{0_{L}} = P(\ket{0_{L}}\ket{\emptyset}\rightarrow\ket{G}\ket{\leftsquigarrow}) = I_{L}$, and the error in the projective measurement of the state $\ket{1_{L}}$ as $E_{1_{L}} = P(\ket{1_{L}}\ket{\emptyset}\rightarrow\ket{G}\ket{\rightsquigarrow}) = I_{R}$. Since the $\ket{0_L}$ is perfectly chiral, all errors arise from the $\ket{1_L}$ state, which has a small $E_{1_L}=1.92\%$ probability to emit a photon in the incorrect direction. This implies a maximum measurement error of $1.92\%$ over all states in the Bloch sphere.

Instead of using the multi-gate protocol from Eq.~\eqref{Eq:PerfectChiralStates}, there is a simpler projective measurement that only requires one local gate $S_2(\theta)$ acting on the second atom, assigning $\ket{0_L}$ and $\ket{1_L}$ as measurement outcomes when photons are detected on the right and left ports. The protocol is inspired by the fact that perfect chiral single-photon emission at  $t=0$
\begin{equation}
	c_{1}(0)+e^{ \mp i\phi}c_{2}(0) = 0, ~ \mathrm{and} ~ c_{1}(0)+e^{ \pm i\phi}c_{2}(0) \neq 0, \label{Eq.Cond1}
\end{equation}
may be achieved if $c_1(0)=e^{i\theta}c_2(0)$ with $\theta \pm \phi = \pi$, to selectively achieve $|\alpha_{L}(0)| = 0$ and $|\alpha_{R}(0)| \neq 0$. While this condition strictly works only at $t=0$, using the same parameters as before $(J,\Delta_\text{ch},\phi_\text{ch})\simeq(0.1\gamma_0,-0.132\gamma_0,0.088\pi)$, we achieve a rather good chirality on both qubit states at all times (cf. Fig.~\SubFig{Fig:MeasuError}{b,c}). An optimization of the experimental parameters allows us to achieve, by deviating slightly from this condition $(\Delta,\phi,\theta) \simeq (-0.1293\gamma_{0},0.0266\pi,0.9974\pi)$, chiralities for the zero and one states, $\eta_{\ket{0_L}} \approx 0.976$ and $\eta_{\ket{1_L}} \approx -0.975$, on par with the best experimental chirality in superconducting platforms~\cite{kannan:2023}, leading to a maximum measurement error of $1.27\%$ over the Bloch sphere.

Both protocols are robust in the neighborhood of the optimal operation point $(\Delta_\text{ch},\phi_\text{ch})$. The errors in the simple projective measurement, subject to deviations in the third atom's detuning $\Delta = \Delta_{\mathrm{ch}} + \delta \Delta$ and in the relative phase acting on the second atom $\theta = \theta_{\mathrm{ch}} + \delta \theta$, for fixed $\phi$, are shown in Figs.~\SubFig{Fig:MeasuError}{d} and~\SubFig{Fig:MeasuError}{e}. The measurement errors are more sensitive to imperfect control of the detuning parameter $\Delta$, and more robust to deviations in $\theta$. In both cases, however, the errors increase by only a few percent for relative changes between 5\% and 10\% in both quantities.


\paragraph{Conclusions.}

We have shown that a triatomic giant molecule in waveguide QED simultaneously supports a fully operational decoherence-free (DF) qubit, protected against radiative loss into the waveguide, and state-dependent chiral single-photon emission. Both properties arise from the same physical mechanism: photon interference at the molecule's coupling points, tuned by the accumulated phase $\phi$. The qubit is encoded in localized W-states and controlled entirely through local operations, with initialization from a single atomic excitation, universal gates through a single tunable frequency (or controlled qubit-qubit interactions), and readout via directional photon emission. The unification of decoherence protection and chiral readout in a single minimal device, governed by one tunable parameter, distinguishes this approach from prior DF schemes that address these capabilities separately.

Operating in the single-excitation subspace, this system is a natural building block for modular quantum computation in waveguide QED~\cite{jiang:2007,monroe:2014,leung:2019,niu:2023,croot:2025}. Recent experiments have demonstrated chiral quantum interconnects between superconducting modules using a similar giant-atom platform~\cite{almanakly:2025}. The present work adds decoherence protection to this architecture, pointing toward modular networks where qubit coherence is preserved against decay into the very waveguide channels that connect the nodes.

As future work, it remains to explore non-local arrangements for implementing qubit-qubit operations in the waveguide, where inspiration may be drawn from the giant atom community~\cite{soro:2023,chen:2026,chen:2025}.
Beyond the chiral readout demonstrated here, the molecule supports a complementary regime at $\phi=0~(\mathrm{mod}~2\pi)$ where only the logical subspace couples to the waveguide with uniform strength $\hbar\gamma_0/2$. This capacity to selectively couple and decouple the qubit subspace not only enables qubit-photon entanglement and quantum state transfer---the essential operations for inter-node links in modular networks---but also opens further prospects in chiral quantum optics~\cite{lodahl:2017} and many-body physics~\cite{fayard:2021,cardenas-lopez:2023} with complex emitters in waveguide QED, including topological setups~\cite{bello:2019,amgain:2024}.

\begin{acknowledgments}
	\paragraph{Acknowledgments.} The authors thank Dr. Alejandro Gonzalez-Tudela and Tomás Levy-Yeyati (IFF-CSIC), and Dr. Lei Du (Chalmers University) for useful discussions.
	YW acknowledges support by the Chinese Scholarship Council (CSC) under the Grant No. CSC202406140039.
	JJGR acknowledges support from CSIC's JAE Chair 2024 program and CSIC's PRO-ERC AGAIN 2025 funding.
	ACS acknowledges financial support from the Comunidad de Madrid through the program Talento 2024 `César Nombela', Grant No. 2024-T1/COM-31530 (Project SWiQL).
\end{acknowledgments}

\emph{Data Availability.---} The code used to generate all theoretical data presented in this letter is available from the authors upon reasonable request.

\bibliography{bib-list.bib}

\end{document}